\documentclass{article}
\include{epsf} 

\begin{document}
\title{\Large \bf Statistical Dynamics of Religions and Adherents}\author{ \large \bf M. Ausloos and F. Petroni \\ SUPRATECS, B5, Sart Tilman, \\ B-4000 Li\`ege, Belgium\\ }
\maketitle
\begin{abstract}
Religiosity is one of the most important sociological aspects of
populations.  All religions may evolve in their
beliefs and adapt to the  society developments. A religion is a social
variable, like a language or wealth, to be studied like any other
organizational parameter.

Several questions can be raised, as considered in this study: e.g. (i)
from a ``macroscopic'' point of view :  How many religions exist at a
given time? (ii) from a ``microscopic'' view point: How many 
adherents belong to one religion?  Does the number of adherents
increase or not, and how?  No need to say that if quantitative answers and
mathematical laws are found, agent based models can be imagined to
describe such non-equilibrium processes.

It is found that empirical laws can be deduced and related to preferential
attachment processes, like on evolving network;  we propose two different algorithmic  models reproducing as well the data. Moreover, a
population growth-death equation is shown  to be a plausible modeling of
evolution dynamics in a continuous time framework. Differences with language dynamic competition is
emphasized.

\end{abstract}

\section{Introduction}
All features of societies (beliefs, attitudes, behaviors, languages,
wealth, etc.) are due to competition \cite{Ax97}. Recently
 the dynamics of world's languages,
especially on their disappearing due to competition with other languages
\cite{Abr+03} has been of interest. It is fair to examine whether such considerations can be applied to religions.

We do not enter into any discussion on the definition of a religion; we
recognize that there are various denominations which can impair data
gathering and subsequent analysis; like many, we admit to put on the same
footing religions, philosophies, sects and rituals. Idem for adherents or
adepts;  there are also agnostics, atheists or ``not concerned''. In fact,
a similar set of considerations exists when discussing languages and
dialects, slangs, etc. Indeed it is expected that there are many
similarities, although many are differences\footnote{If it is possible to be bilingual, it is not common to be ``bireligious''}, between the diffusion,
relaxation and distribution of languages and religions.   What is their
geographical distribution? What is their life time? How do they evolve, -
from monotheism to polytheism and ``backwards''? How long does an
adept/adherent remain in one religion? Moreover, even though many
societies are thought to form a hierarchy, due to a competition between
individual concerns, as explained by  Bonabeau et al. \cite{Bonab95} or
discussed by Sousa and Stauffer \cite{SoSt00}, such considerations for
religion should be left for further investigation. These questions need
much more reliable data than it seems available and practical at this
time. Thus,  let us claim that we are not interested here in religion's
origin, activity,  history or hierarchy, but rather in statistical physics
aspects of a non-equilibrium agent based system. We will then consider as parameters the numbers of adherents of each religion, and only these numbers will be treated as physics object (and not the religions themselves).

To address these issues, we have followed classical scientific steps as in
physics investigations. We have analyzed ``empirical'' data on the number
of adherents of religions taken from two different freely available data
sets.

The next scientific step is to analyze the data along statistical physics
modern lines. Zipf and Pareto-like plots will be given. After deducing
empirical laws, a theoretical modeling is in order. In view of the
observed features, and following standard intuition, one thinks at once
about two algorithmic ``agent based'' models,  describing preferential
attachment on a network, as in the Potts model \cite{Potts} of magnetism
or  Axelrod model \cite{Ax97} in sociology, already applied in opinion
formation studies \cite{holyst}.

Thereafter studying the time evolution of several ``main'' religions, we
observe that a microscopic interpretation is plausible along the lines of
a growth Avrami equation in a continuous time framework. This equation
seems more plausible than the generalized Verhulst equation for modeling
the  dynamics of language deaths \cite{Abr+03} because the former allows
better handling of ``internal and/or external fields'' (as those mentioned
above, - most of them missing in language dynamics) as well as
(microscopic) homogeneous and/or heterogeneous fluctuations at an early
stage of evolution, - while the Verhulst equation of  Abrams and Strogatz
\cite{Abr+03} is $grossly$ empirical. Notice that
languages were simulated also with other than Lotka-Verhulst-Volterra
mechanisms \cite{Abr+03};
see e.g. ref. \cite{viviane}.

\section{Data}
 The first data set is taken from The International Data Base
(IDB)\cite{footnote1}.
Data on Religions are included in table 58 and contains information on the
population of 103 nations worldwide. The surveys were carried between 1960
and 1992. In the dataset are recorded the number of adherents of 150
religions, taking into account about 2 billion people ($1/3$ of the
present world population).

The second data set was taken from the World Christian Encyclopedia (WCE)
\cite{WCE}, it gives information on the number of adherents of the world's
main religions and their main denominations (56 religions overall)
considering the whole world population. From this data set we have also
information on changes during one century of the number of adherents of
each religion from 1900 till 2000, measured over a 5 year span,  with a
forecast for 2025 and 2050.  No need to say that further work should go
back to history: the number of ``religions'' is highly time dependent,
the more so when one distinguishes them to the level of denominations and
sects; the number of adherents of a given religion is not fixed either.
History is full of examples of individuals or entire groups of people
changing their religion, - for various reasons: following the ``leader''
(e.g. Constantinus, ...) or ``external pressure'' (e.g.
inquisition, ...) or ``internal pressure'' or so
called adaptation under proselytism action...

One should also be aware that such surveys are biased, and are hardly
snapshots of a situation like in a laboratory experiment. Yet, beside
these caveats, the main difference between the two data sets is in the
information they give on religions with a small number of adherents. While
this information is present (even if not for all considered nations, and
only partially) in the first data set, the second data set does not
consider small religious groups.  It is also unclear how much distinction
was made in the IDB and WCE surveys concerning denominations and sects so
called $adstrated$ to the main religions.

\section{Zipf's and Pareto's distributions}

\begin{figure}[ht]
\begin{center}
$\begin{array}{cc}
\multicolumn{1}{l}{\mbox{\bf (a)}} &	
\multicolumn{1}{l}{\mbox{\bf (c)}} \\ 
\epsfxsize=4.0truecm\epsffile{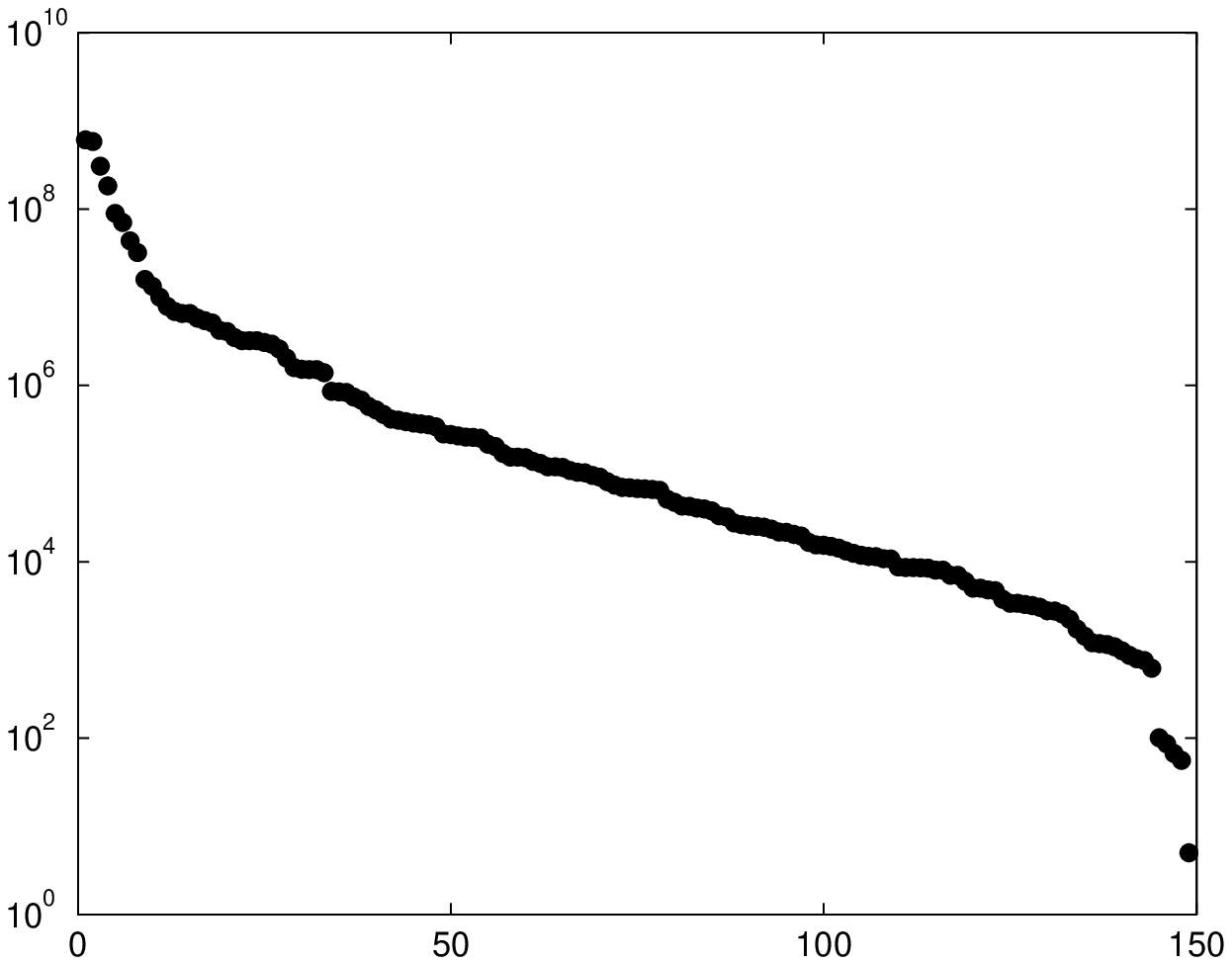} &	
\epsfxsize=4.0truecm	\epsffile{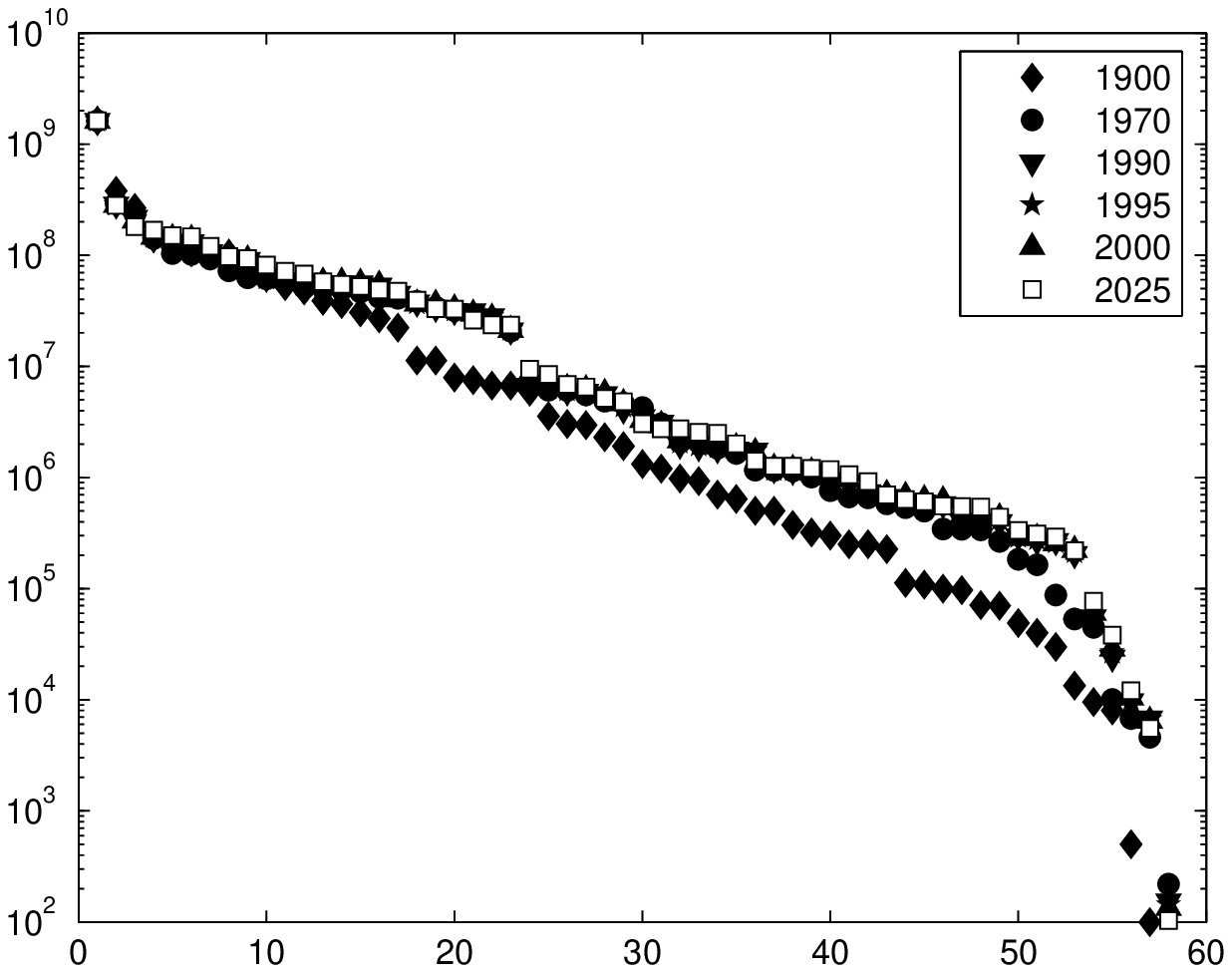} \\	\multicolumn{1}{l}{\mbox{\bf (b)}} &	
\multicolumn{1}{l}{\mbox{\bf (d)}} \\ 	\epsfxsize=4.0truecm\epsffile{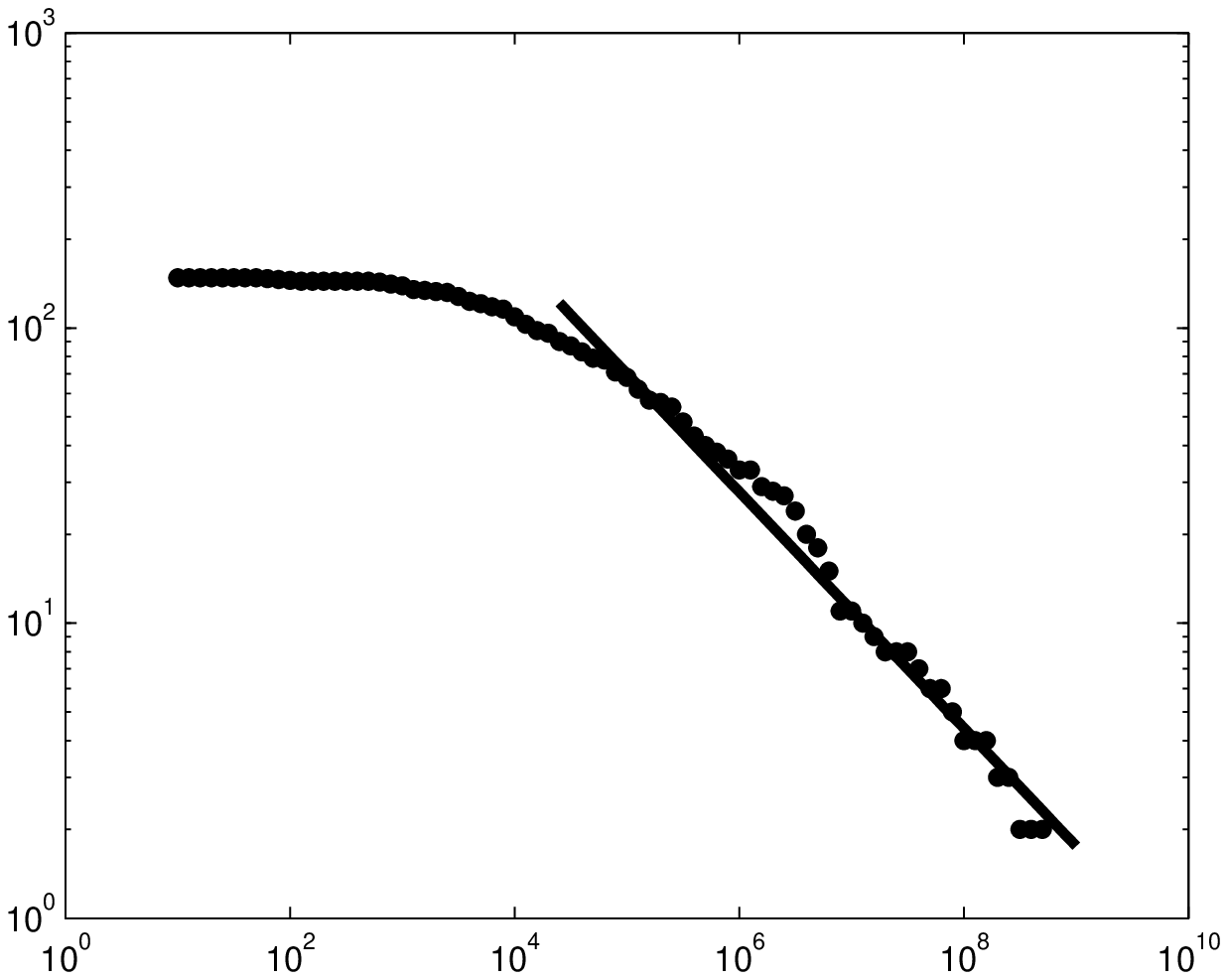} &	
\epsfxsize=4.0truecm	\epsffile{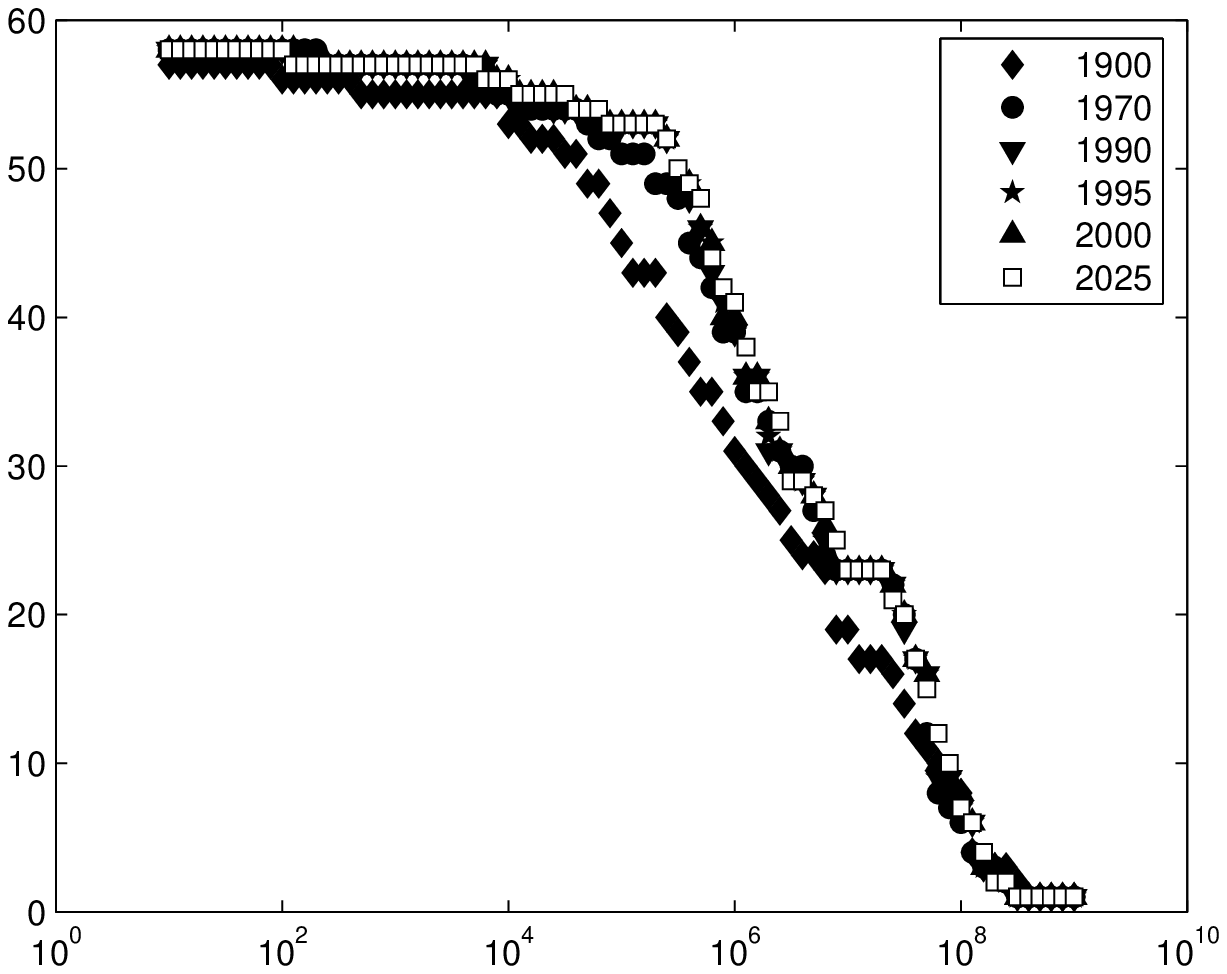}\\	[0.4cm]
\end{array}$ 
\end{center}
\caption{Zipf's and Pareto's distributions of religions. Subplots {\bf (a)} and {\bf (c)} show the Zipf's distribution for the IDB and WCE data sets respectively. On the $y$ axis is the number of adherents;  on the $x$ axis the ranked religions.Subplots {\bf (b)} and {\bf (d)} show the Pareto distributions for these data sets. These plots show the number of religions ($y$ axis) with a number of adherents $n>N$ as function of $N$. The axis scales have been chosen to enlighten linear regions}\label{fig1}
\end{figure}

The Zipf's and Pareto's distributions are shown in figure \ref{fig1}  for
both data sets. Recall that the Zipf distribution results from a
hierarchical ranking (of the considered religions according to their
number of adherents). The Pareto distribution shows instead the number of
religions with a number of adherents $n$ greater than $N$ as a function of
$N$. In figure \ref{fig1}(a) and \ref{fig1}(b), the Zipf and Pareto
distributions are shown respectively for the first dataset while
\ref{fig1}(c) and \ref{fig1}(d) show results for the second data set in
different (so-called) years. It can be noticed that the Zipf distribution
for both data sets can be fitted by a straight line, with different
slopes,  - except for the tails, i.e. where religions with a very small or
high number of adherents are to be found (see $caveats$ above).   However
it is $remarkable$ that a different behavior between both data sets is
found in the case of the Pareto distribution: while for the IDB data set,
Fig.\ref{fig1}(b), it can be seen that the Pareto distribution roughly follows a
power law at least for $N>10^5$, i.e. $f(N)\propto N^{-0.4}$; this is
not the case for the WCE data set, Fig. \ref{fig1}(d), where the linearity
is present only in a log-linear plot. Notice that the former law exponent of the Pareto distribution  is similar to that found in language studies \cite{viviane}. Such an
empirical non trivial power law is consistent with a preferential
attachment process \cite{prefatt} on a network, i.e. it is more likely
that one has the religion of one's mother or neighbor....

\section{Partial distribution functions}
\begin{figure}\epsfxsize=11.0truecm
\epsffile{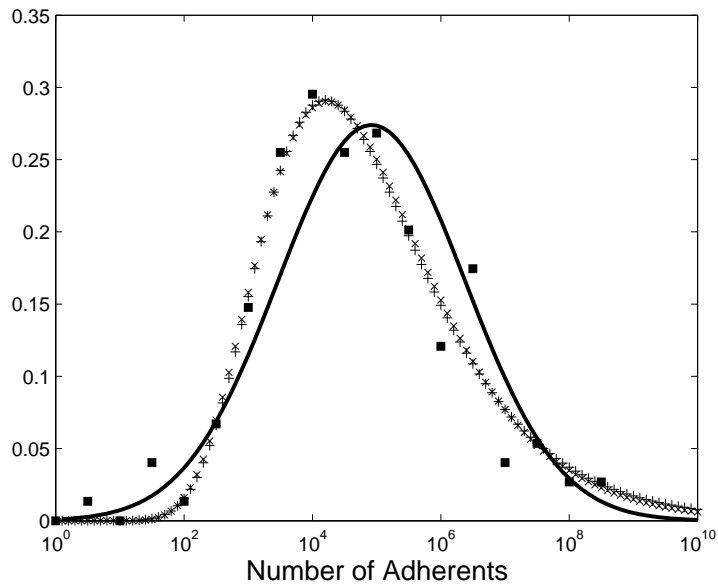}
\caption{Partial Distribution Function (pdf) of adherents. The distribution of the number of adherents of religions from the IDB dataset is shown (squares);  an exponentially increasing bin size is used for the $x$-axis. The pdf is fitted with Weibull (+) or log-normal (x) distributions and compared with the best Gaussian fit (continuous line).}\label{fig2}
\end{figure}

In order to compare the two data sets, and their meaning or bias,  and observe the time evolution of
adherence (or attachment) we have divided the population interval
$[1,10^9]$ into 18 bins of exponentially increasing size and filled each
bin with the number of religions having that number of adherents
(normalized to have the distribution area equal to 1). The result is a
partial distribution function (pdf), Fig.\ref{fig2}, that can be fitted (i)
with a Weibull distribution (symbol +), much used in life time (or failure) studies,   
\begin{equation}\label{weibull}
f(x)={1\over \beta} e^{-{(x-\mu)\over \beta}} e^{-e^{-{(x-\mu)\over
\beta}}} \end{equation}
where $x=\log_{10}(n)$ and $n$ is the number of adherents $or/and$ (ii) with a lognormal
distribution (symbol x); both fits are quite similar, with a slight
difference in the upper tail. For comparison the best corresponding
Gaussian distribution (continuous line) is shown in the same plot. This
leads to consider that  $two$ empirical functions can be possible based on different concepts at this level
of data acquisition : (i) birth-death processes\footnote{We realize that $x$ is Eq.(1) is the size of the population, while the variable of the Weibull distribution is
rather the strength of to-be-broken bonds in a ``time to failure''
analysis. If there is a one-to-one correspondence between the $x$ and $y$
axes in cause-effect relations, such a change in meaning is only a change
in notations. Otherwise, hysteresis effects are to be considered. This
goes beyond our present study.},  (ii)
multiplicative production with many  independent factors.

\begin{figure}\epsfxsize=11.0truecm
\epsffile{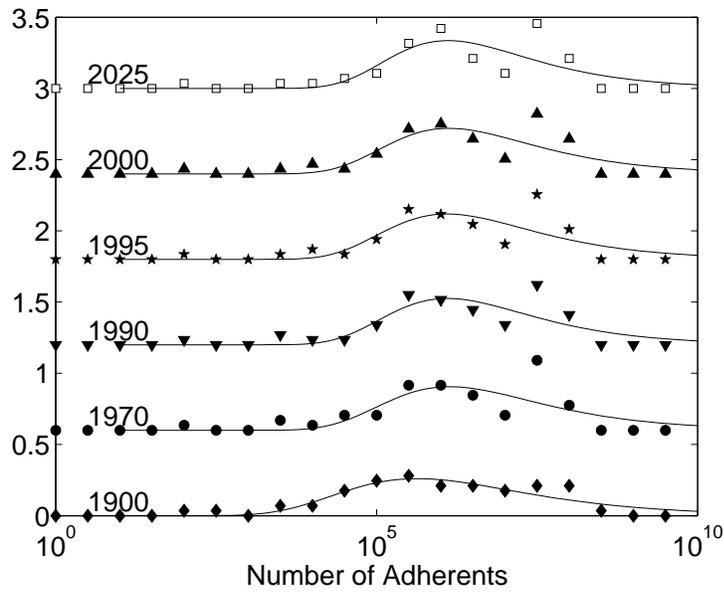}
\caption{Time evolution of Partial Distribution Functions of religion sizes. The distribution of the number of adherents of religions from WCE data set is shown according to an exponentially increasing bin size on the $x$-axis. Results for different ``years'' are vertically displaced of $0.6$ in order to have them on the same plot. The fit is done using a Weibull distribution (continuous lines).}\label{fig3}
\end{figure}

The same procedure can be  applied to the WCE data set, whence obtaining
the pdf's shown in Fig. \ref{fig3} for different  (so called) years. To
eliminate the effect due to increasing world population in the ``years''
considered, all pdf's of different ``years'' were normalized to the same
population number considering 1900 as the reference population. A fit of
these distributions, with  Eq.(\ref{weibull}), is shown in  Fig.\ref{fig3}. In order to plot all the pdf's on the same graph each pdf has
been successively displaced by $0.6$.   The apparent flatness of the pdf
is due to the vertical rescaling. From this figure a critical view of this
data has to be implied: notice the break at $10^7$, indicating in our view
an overestimation of adepts/adherents  in the most prominent religions, or a lack of
distinctions between denominations, as can be easily understood
\cite{morelli}. This ``emphasis'' of the ``winner takes all'' in the WCE data, i.e. the
empirical data results from summing up adherents from the most important
sort of religion and  smaller related denominations into a single number
corresponding to (the main) religion, hints to explaining the difference between Pareto plots in Figs. \ref{fig1}(b), - \ref{fig1}(d), 

\section{Time evolution}
Finally, it is easily accepted that the percentages of adherents are not
fixed over time. Therefore a nucleation-growth-death process can be
proposed, in analogy with crystal growth studies
\cite{auslooscrystalgrowth}. We consider that a microscopic-like,
continuous time differential equation can be written for the evolution of
the number of adherents (in terms of percentage with respect to the world
population) of the world main religions, as for competing entities of the
type \cite{Gadom}
\begin{equation}\label{Avrami}
{d \over dt} g(t)=S k(t)[1-g(t)] {dV_n \over dt}
\end{equation}
where, adapting to our case this Avrami-Kolmogorov
equation, $g(t)$ is counting the fraction of adherents of a given religion, $V_n$ is instead connected with the total world population, $S$ is a parameter
to be determined and $k(t)\propto t^{-h}$ where $h$ is a parameter  to be
deduced in each case, measuring the attachment-growth (or death) process
in this continuous time approximation. This should be contrasted with the
Lotka-Volterra-Verhulst mechanistic approach (for languages) which hardly allows for
nucleation, dissipation and/or time delayed correlations of different
entities, in contrast to generalizations of   Eq. (2) using such physical features.

\begin{figure}\epsfxsize=11.0truecm
\epsffile{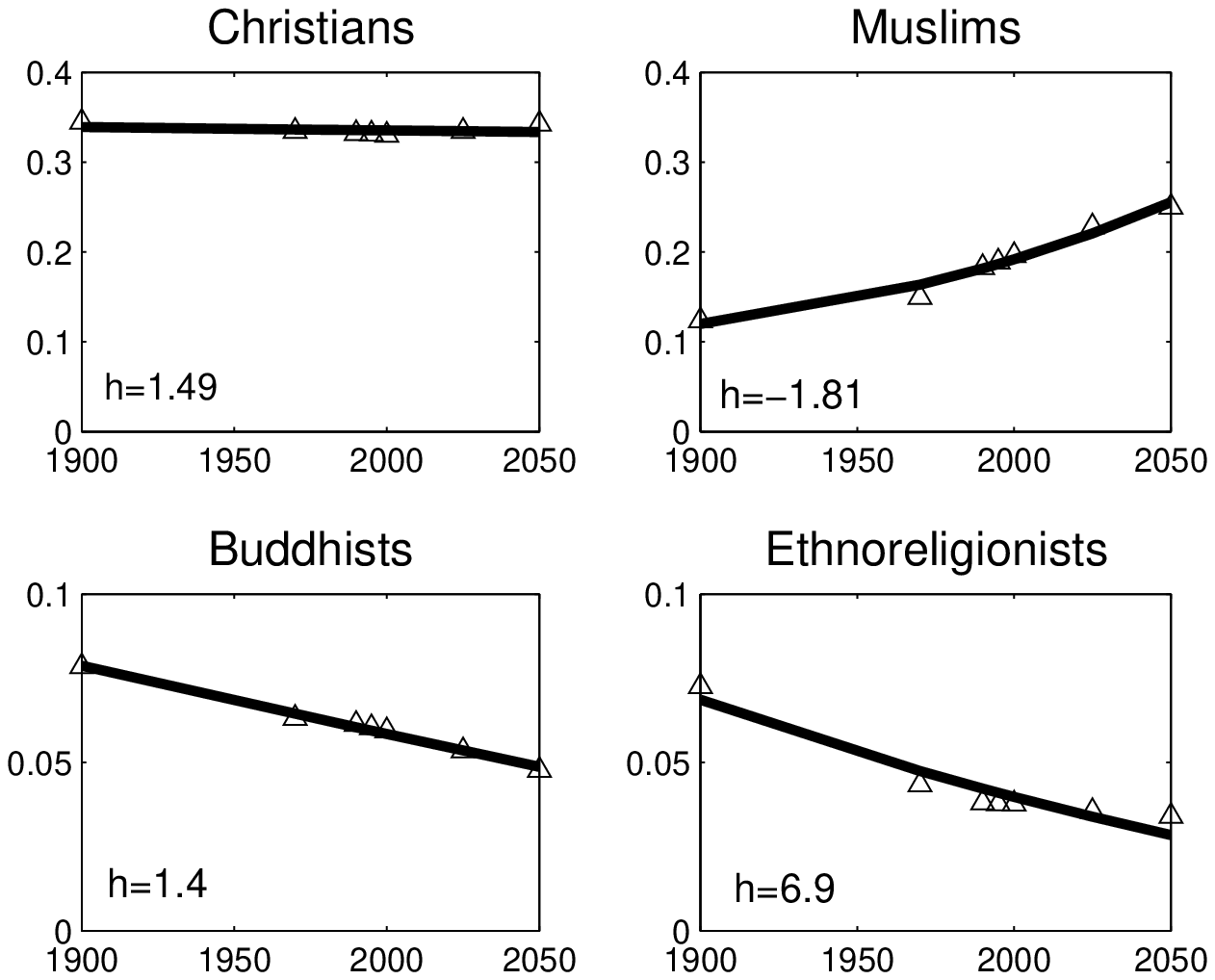}\caption{Time evolution of adherents from the WCE data set. The plot shows
the percentage of adherents for 4 typical world religions as a function of
time. Each value of the attachment parameter $h$ as given by the best fit is reported in the plots}\label{fig4}
\end{figure}

A few examples of religions for which the number of adherents is
increasing (e.g., Islam), decaying (e.g., Ethnoreligions and Buddhists) or
rather stable (e.g., Christianity) is shown in Fig.\ref{fig4}. The data
can be well fitted to the solution of the Avrami-Kolmogorov
growth-death equation Eq.(2). The values of $h$ for the considered religions, as obtained by a least-square best fit, are reported in the plot. The parameter $h$ values and their meaning deserve some short explanation and discussion.  The parameter can be thought to be like a
reproduction rate in Verhulst logistic equation, or a true attachment like
in sexual networks \cite{sex} or in molecular processes \cite{mole}. It is
interesting to observe that $h$ can be positive or negative, indicating
also the possibility for $detachment$.  Other parametrizations of $k(t)$
can be imagined and are possible. Our theoretical law elsewhere derived
from first principles \cite{Gadom} concludes the present scientific analysis in
showing that a predictability level can be reached on the evolutions.

\section{Conclusions}
In conclusion, as for languages or wealth, one can recognize religions as
a signature of population dynamics. Even though characteristic time scales
are different, and religion dynamics is more complex than language
dynamics because of the presence of external fields and spontaneous
nucleations, empirical ranking laws look similar. Therefore similar
growth-death agent based models can be thought of. Yet, there are useful
differences to be expected (and found) which lead to different models from
those describing language death and appearance. We  propose an
algorithmic approach based on attachment processes for the macroscopic
point of view, - not deciding on the statistical alternative, i.e. Weibull or log-normal law, and a diffusion growth rate based equation for modeling the
data at the microscopic level. There are possible open problems on the ongoing research, or further investigations taking into account the available/reliable data at this time, as to look for (time dependent) geographical effects, like clustering, or through other definitions, like normalizing with respect to some population size or country surface, or
GDP, or other socio-economic index allowing to build correlation matrices
and search for socio-economic field influence.

\vskip 1.6cm {\large \bf Acknowledgments} \vskip 0.6cm

The work by FP has been supported by European Commission Project
E2C2 FP6-2003-NEST-Path-012975  Extreme Events: Causes and Consequences.
Critical and encouraging comments by  A. Morelli have been very valuable.
Referees should be thanked, moreover for their warning and putting
pressure on us to emphasize that we mere treat religions and adherents as
physics variables, so that our results and their interpretation have never the intention of vilifying any religion, sect, person, etc.

\newpage

{\large \bf Figure Captions}

\vskip 0.5cm{\bf Figure 1} --  Zipf's and Pareto's distributions. Subplots
{\bf (a)} and {\bf (c)} show the Zipf's distribution for the IDB and WCE
data sets respectively. On the $y$ axis is the number of adherents;  on
the $x$ axis the ranked religions. 
Subplots {\bf (b)} and {\bf (d)} show the Pareto distributions for these
data sets. These plots show  the number of religions ($y$ axis) with a
number of adherents $n>N$ as function of $N$. The axis scales have been chosen to enlighten linear regions

\vskip 0.5cm {\bf Figure 2} --  Partial Distribution Function (pdf) of
adherents. The distribution of the number of adherents of religions from
the IDB dataset is shown (squares);  an exponentially increasing bin size
is used for the $x$-axis. The pdf is fitted with Weibull (+) or lognormal
(x) distributions and compared with the best Gaussian fit (continuous
line).

\vskip 0.5cm{\bf Figure 3} -- Time evolution of Partial Distribution
Functions of religion sizes. The distribution of the number of adherents
of religions from WCE data set is shown according to an exponentially
increasing bin size on the $x$-axis. Results for different ``years'' are
vertically displaced of $0.6$ in order to have them on the same plot. The
fit is done using a Weibull distribution (continuous lines).

\vskip 0.5cm{\bf Figure 4} -- Time evolution of adherents from the WCE data set. The plot shows
the percentage of adherents for 4 typical world religions as a function of
time. Each value of the attachment parameter $h$ as given by the best fit is reported in the plots.

\newpage

\newpage

\newpage

\newpage



\begin{thebibliography}{99}
\bibitem{footnote1}The International Data Base (IDB) is a computerized
source of demographic and socioeconomic statistics for 227 countries and
areas of the world. The IDB provides a quick access to specialized
information, with emphasis on demographic measures, for individual
countries or selected groups of countries in the world. The major types of
data available in the IDB include:
Population by age and sex, Vital rates, infant mortality, and life tables,
Fertility and child survivorship, Migration,    Marital status, Family
planning,       Ethnicity, religion, and language, Literacy, Labor force,
employment, and income, Households. Sources of the data include:        U.S.
Census Bureau, Estimates and Projections,       National Statistics
Offices,        United Nations and Specialized Agencies (ILO, UNESCO, WHO))
\end{thebibliography}
\end{document}